# Polarized Skylight Navigation Simulation (PSNS) Dataset

Huaju Liang, and Hongyang Bai

*Abstract*—**With more and more machine learning methods applied to bioinspired polarized skylight navigation, the demand for polarized skylight navigation datasets is more and more urgent, which can be used to train and test machine learning methods. So, in this paper, an open polarized skylight navigation dataset is constructed for the first time. Firstly, a polarized sky model was proposed based on Sun position model, Berry sky model and Hosek sky model, which contains the information of the light intensity (LI), degree of polarization (DOP) and angle of polarization (AOP). Secondly, a polarization imaging simulation system is constructed, which can capture not only LI, DOP and AOP images, but also original black-and-white LI images in different polarization directions. Black-and-white LI images are the original data collected by actual polarization imager, so this system can completely describe the whole process of polarization imager capturing skylight polarization patterns. Above all, a polarized skylight navigation simulation (PSNS) dataset can be constructed. In addition, to facilitate researchers to build their own datasets based on their own polarized light sensors and sky models, we have disclosed the source code of polarization imager and original LI imager on GitHub.**

## I. INTRODUCTION

The atmospheric scattering of sunlight from gas molecules, aerosols and dust causes the skylight to have a polarization pattern. According to this pattern, new generation aerosol and the density of PM2.5 (atmospheric particulate matter that has a diameter of less than 2.5 micrometers) particulates are estimated usually based on polarized sensor [1-3]. In addition, the complementary information of polarimetric brings remarkable improvements in remote sensing image classification and target recognition [4, 5]. Moreover, recently, it has been reported that many insects can use skylight polarization pattern trails to find their way [6-8]. For instance, desert ants determine heading with great efficiency using polarization compass cues [9, 10]; Dung beetles perform navigation tasks relied on the dorsal rim area (DRA) of compound eyes, which is sensitive to polarized light [11, 12];

Drosophila sense the angle of polarization (AOP) in the sky as an orientation cue [13, 14].

Inspired by insects, polarized skylight navigation has been applied in robot navigation [7], autonomous unmanned aerial vehicle navigation [15], smart-munition platform navigation [16] and even underwater navigation [17, 18]. And many traditional polarized skylight navigation methods have been proposed, such as zenith method [19, 20], symmetry detection method [21, 22], least square method [23-25], gradient method [16] and solar meridian and anti-solar meridian (SM-ASM) method [26, 27]. In addition, machine learning methods are gradually applied to polarized skylight navigation [28, 29]. With machine learning methods applied to polarized skylight navigation, polarized skylight navigation datasets are urgently needed to train and test these machine learning methods. So, a skylight polarization simulation system is designed in this paper to construct an open polarized skylight navigation simulation (PSNS) dataset for the first time.

Skylight polarization pattern as a result of scattering phenomena is well described by Rayleigh scattering theory [30, 31]. Because Rayleigh single-scatter sky model is simple and practical, it is widely used in skylight polarization navigation, remote sensing and so on [7, 32-35]. However, Rayleigh single-scatter sky model has three basic assumptions: 1. The scattering particles are smaller than the wavelength of incident sunlight; 2. The scattering particles are isotropic, homogeneous spheres; 3. The sunlight only undergoes a single scattering event [31, 34, 36]. So, the skylight polarization patterns prescribed by the Rayleigh theory have limitations. Monte Carlo sky model can well describe multiple scattering through Monte Carlo experiments, but it is not an analytical model [37, 38]. According to the Rayleigh single-scatter sky model, there are only two unpolarized points (sun and anti-sun points) in the sky. In practical observation, however, there are four unpolarized points in the sky [39], which are named neutral points.

According to the phenomena of four neutral points in the sky, M. V. Berry has proposed the Berry model [40, 41], which is an analytical model. The Berry model can describe the neutral points by introducing singularities and the model parameters can be modified according to the actual measurements [42-44]. So, the Berry model is close to practical atmospheric polarization patterns and it is an analytical model with low computational cost and contains the information of AOP and degree of polarization (DOP) [40, 43]. However, the Berry

This work was supported in part by the National Natural Science Foundation of China under Grant 61603189. (Corresponding author: Hongyang Bai.)

H. Liang and H. Bai are with the School of Energy and Power Engineering, Nanjing University of Science and Technology (NJUST), Nanjing 210094, China (e-mail: lianghuaju@sina.cn ; hongyang@njust.edu.cn)

model overlooks the information of light intensity (LI). To address this problem, in this paper, Hosek sky model is introduced to describe the luminance of clear sky [45, 46]. And this model can significantly correct the irradiance of sky and tends to best fit the trend of measured irradiance curves [47]. So, the combination of Berry and Hosek models can fully describe skylight polarization patterns.

In addition, a hypothetical polarization imager has been established to promote the practical application of skylight polarization patterns. According to Rayleigh sky model, several hypothetical polarization imagers are constructed to capture the AOP image and DOP image [16, 27, 28, 48], which facilitate the verification of polarized skylight navigation and polarization remote sensing methods. However, the original data collected by the polarization imager are black-and-white LI images in different polarization directions. Polarization imagers are mainly divided into rotating polarizer imager, multi-channel polarization imager, and pixelated polarization imager. Rotating polarizer imager captures black-and-white LI images in different polarization directions by rotating polarizer [49-52]. Multi-channel polarization imager uses several image sensors with polarizers in different polarization directions to capture black-and-white LI images in different polarization directions [53-55]. Pixelated polarization imager based on a four-directional polarizer installed on pixels to capture black-and-white LI images in different polarization directions [8, 56, 57]. Then, these polarization imagers fuse several black-and-white LI images in different polarization directions to capture AOP and DOP images. So, capturing AOP and DOP images directly cannot fully reflect the whole process of polarization image acquisition. Aiming at this problem, a polarization imaging simulation system is constructed, which can capture not only AOP and DOP images, but also the original black-and-white LI images in different polarization directions. Finally, adding measurement noises to the original black-and-white LI image, the PSNS dataset is constructed.

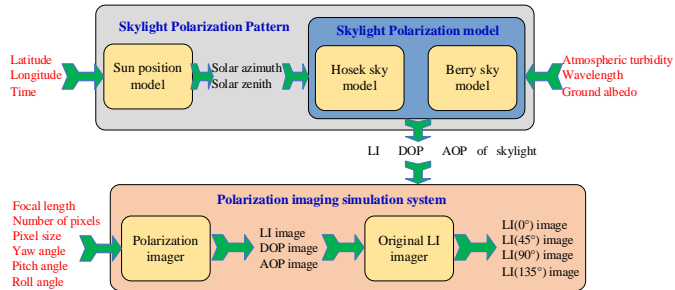

Fig. 1. Structure of skylight polarization simulation system and dataset building process, where LI represents light intensity, DOP represents the degree of polarization, AOP represents the angle of polarization, $LI(0°)$, $LI(45°)$, $LI(90°)$ and $LI(135°)$ images represent the original black-and-white LI images in $0°$, $45°$, $90°$, and $135°$ polarization directions.

The skylight polarization simulation system and PSNS dataset building process are shown in Fig. 1. Firstly, solar azimuth and zenith angles are determined by Sun position model. Secondly, skylight polarization model is constructed based on Berry and Hosek sky models, so this model includes the information of LI, DOP and AOP. Thirdly, a polarization imaging simulation system is constructed based on the designed skylight polarization model to capture not only AOP and DOP images, but also original black-and-white LI images in different polarization directions to construct PSNS dataset.

## II. SKYLIGHT POLARIZATION PATTERN MODELING

In this section, a skylight polarization model is constructed based on Sun position model, Hosek sky model, Berry sky model and Neutral point model, which fully describes skylight polarization patterns by the information of LI, DOP and AOP.

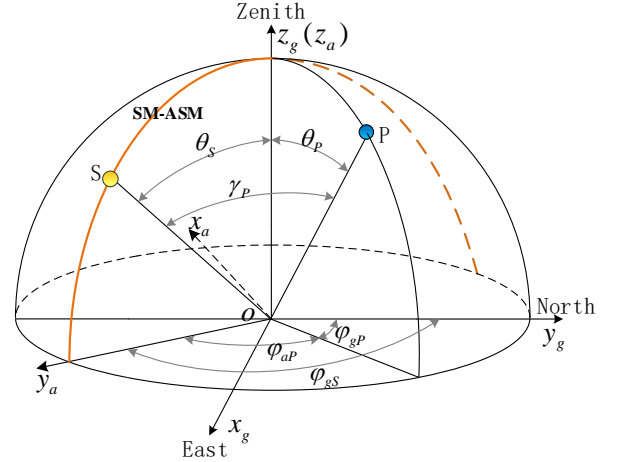

Fig. 2. Cartesian coordinate systems, where $ox_g y_g z_g$ is the East-North-Up (ENU) Cartesian coordinate system and $ox_a y_a z_a$ is the solar azimuth Cartesian coordinate system. The yellow point S and blue point P represent the Sun and observation point, respectively. The orange line SM-ASM is the solar meridian and anti-solar meridian. $\theta_S$ and $\theta_P$ represent the zenith angles of S and P, respectively. $\gamma_P$ represents the angle between S and P, which is named scattering angle. $\varphi_{gS}$ and $\varphi_{gP}$ represent the azimuth angles of S and P in the ENU coordinate system, respectively. $\varphi_{aP}$ represents the azimuth angle of P in the solar azimuth coordinate system.

To well describe skylight polarization patterns, we construct a solar azimuth Cartesian coordinate system. As illustrated in Fig. 2, $ox_g y_g z_g$ is the East-North-Up (ENU) Cartesian coordinate system. The origin of solar azimuth Cartesian coordinate is $o$, the direction of solar azimuth is labeled by $y_a$, the direction of solar zenith is labeled by $z_a$, and $x_a$ completes the construction of solar azimuth right-handed Cartesian coordinate system $ox_a y_a z_a$.

### A. Sun position model

Skylight polarization patterns are regular relative to the Sun position. So, the Sun position is given by the formulas of celestial navigation, which can be found in [58-61].

### B. Skylight polarization model

A full description of skylight polarization patterns must contain the information of LI, AOP and DOP.

#### 1) LI

The Hosek model is an analytic formula for describing the luminance of clear skies with low turbidity [45, 46]. And this model significantly corrects the sky's irradiance and tends to best follow measured irradiance curves [47]. The LI distribution can be obtained by using the Hosek model.

$$LI(\theta_P, \gamma_P) = (1 + Ae^{B/(\cos\theta_P + 0.01)}) \times$$

$$(C + De^{E\gamma_P} + F\cos^2\gamma_P + G \cdot \chi(H, \gamma_P) + I \cdot \cos^{1/2}\theta_P) \quad (1)$$

where $\theta_P$ is the zenith angle of the observation point P, $\gamma_P$ is the angle between the observation point P and Sun, $\chi(H, \gamma_P)$ is anisotropic term for luminance peaks around the Sun. $A$, $B$, $C$, $D$, $E$, $F$, $G$, $H$ and $I$ are adjustable coefficients and the functions of atmospheric turbidity, wavelength and ground albedo.

**2) AOP**

According to the phenomena of four neutral points in the sky, Berry sky model was proposed [40-42]. The Berry model can well describe the characteristics of atmospheric polarization patterns.

$$w(\zeta) = -\frac{4(\zeta - \zeta_+)(\zeta - \zeta_-)(\zeta + 1/\zeta_+^*)(\zeta + 1/\zeta_-^*)}{(1 + |\zeta|^2)^2 |\zeta_+ + 1/\zeta_+^*||\zeta_- + 1/\zeta_-^*|} \quad (2)$$

where $\zeta$, $\zeta_+$, $\zeta_-$, $-1/\zeta_+^*$ and $-1/\zeta_-^*$ are complex numbers in the complex coordinate system, as illustrated in Fig. 3. $\zeta$ represents the projection of observation point on the complex coordinate system. $\zeta_+$, $\zeta_-$, $-1/\zeta_+^*$ and $-1/\zeta_-^*$ represent the projections of four neutral points in the sky. $|\cdot|$ represents the modulus of a complex number.

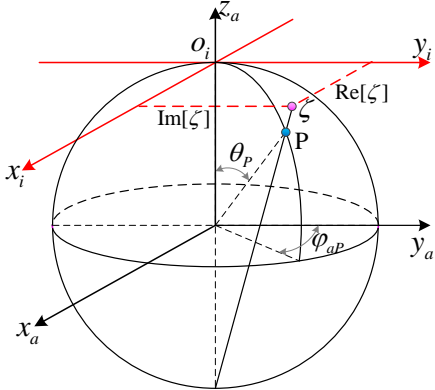

Fig. 3. Complex coordinate system, where $o_i x_i y_i$ is the complex coordinate system and $\omega x_a y_a z_a$ is the solar azimuth Cartesian coordinate system. $x_i$ is parallel to $x_a$, and $y_i$ is parallel to $y_a$. The blue point P represents the observation point, and pink point $\zeta$ represents the projection of P on the complex coordinate system. $\theta_P$ represents the zenith angles of P. $\varphi_{aP}$ represents the azimuth angle of P in the solar azimuth coordinate system. where $\text{Re}[\zeta]$ and $\text{Im}[\zeta]$ represent the real and imaginary part of the complex number $\zeta$. respectively.

Then, the AOP distribution of skylight polarization pattern is given by

$$AOP_a(\zeta) = \arg[w(\zeta)] \quad (3)$$

where $\arg[\cdot]$ represents the principal argument of a complex number. AOP is the angle between polarization electric field vector (E-vector) and reference direction, so the value of AOP is determined by not only polarization E-vector, but also reference direction. The reference direction of $AOP_a$ is the $x_a$ axis of the solar azimuth coordinate system. When discussing skylight polarization pattern, the local meridian is usually used as the reference direction, so $AOP_m$ is defined whose reference direction is the local meridian.

$$AOP_m(\zeta) = AOP_a(\zeta) - \arctan\left(\frac{\text{Re}[\zeta]}{\text{Im}[\zeta]}\right) \quad (4)$$

where $\text{Re}[\cdot]$ and $\text{Im}[\cdot]$ represent the real and imaginary part of a complex number, respectively.

**3) DOP**

To better describe DOP of skylight, the Berry model was modified by considering the influence of atmospheric turbidity, wavelength and ground albedo [43, 44].

$$DOP(\zeta, \theta_P, \gamma_P, T, \rho, \lambda) = |w(\zeta)| \times$$

$$(\theta_P E(\theta_P, \rho) + (\frac{\pi}{2} - \theta_P) S(\gamma_P, \theta_P, \lambda)) M_{DOP}(\tau) \quad (5)$$

where $M_{DOP}(\tau)$ describes the maximal DOP in the sky, and $\tau$ is the atmospheric turbidity. $E(\theta_P, \rho)$ describes the depolarization effects near the earth ground, and $\rho$ is the ground albedo. $S(\theta_P, \gamma_P, \lambda)$ is the influence of spectral radiant power, and $\lambda$ is the wavelength of light.

The maximal DOP in the sky is

$$M_{DOP}(\tau) = e^{-\frac{\tau}{k_1} + k_2} \quad (6)$$

where $k_1$ and $k_2$ are the parameters, and in our model, $k_1$ and $k_2$ are set as 4 and 0.12 [44].

The depolarization effects near the earth ground is

$$E(\theta_P, \rho) = \cos(\theta_P)^\rho \quad (7)$$

The influence of spectral radiant power is

$$S(\theta_P, \gamma_P, \lambda) = \left(\frac{1}{S_D(\lambda)} - \frac{1}{S_{sun}(\lambda)}\right) \frac{S_{90}(\lambda) S_{sun}(\lambda)}{S_{sun}(\lambda) - S_{90}(\lambda)} \quad (8)$$

where $S_D(\lambda)$ is the spectral radiant power based on Perez model, $S_{sun}(\lambda)$ and $S_{90}(\lambda)$ are the spectral radiant power at the sun and 90 degrees from the sun [43].

**4) Polarization E-vector**

In addition, to facilitate the construction of polarization imaging simulation system, which will be described in section 3, polarization E-vector in solar azimuth coordinate is given by

$$\overrightarrow{E_{aP}} = \begin{pmatrix} \sin AOP_a \\ \cos AOP_a \\ -(\sin AOP_a \sin\varphi_{aP} + \cos AOP_a \cos\varphi_{aP})\tan\theta_P \end{pmatrix} \quad (9)$$

## III. POLARIZATION IMAGING SIMULATION SYSTEM

In this section, a polarization imaging simulation system is constructed, which can capture not only DOP and AOP images, but also original black-and-white LI images in different polarization directions.

## A. Polarization imager

To capture polarization images, the imager-centered coordinate system $o_c x_c y_c z_c$ and pixel coordinate system $o_p x_p y_p$ are constructed, as illustrated in Fig. 4. The origin $o_c$ of $o_c x_c y_c z_c$ coincides with the optical centre of the hypothetical polarization imager, and the $z_c$ axis coincides with the optical axis of the imager. The $o_c x_c y_c$ plane is assumed to be parallel to the image plane of the imager and at a distance $f$ to $o_c$, where $f$ is the focal length of the imager. The $o_p x_p y_p$ plane is assumed to be coplanar with the image plane of the imager, $x_p$ and $y_p$ are parallel to $x_c$ and $y_c$, respectively.

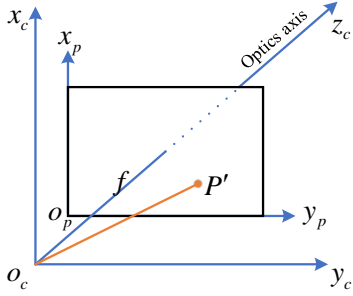

Fig. 4. Imager Cartesian coordinate system $o_c x_c y_c z_c$ and pixel coordinate system $o_p x_p y_p$, where $f$ represents the focal length of the imager, $P'$ represents any point in the $o_p x_p y_p$ plane.

Let $(x_{pP'}, y_{pP'})$ represent the coordinates of any point $P'$ in the pixel coordinate system. Then, the vector corresponding to $P'$ in imager-centered coordinate system is

$$\overrightarrow{o_c P_c'} = \left( D_x (x_{pP'} - \frac{\eta_x + 1}{2}), D_y (y_{pP'} - \frac{\eta_y + 1}{2}), f \right)^T \quad (10)$$

where $\eta_x$ and $\eta_y$ represent the image plane has $\eta_x \times \eta_y$ pixels, $D_x$ and $D_y$ represent the pixel sizes in $x_p$ and $y_p$ directions, respectively.

The rotation matrix from the imager-centered coordinate system to the solar azimuth coordinate system is

$$C_c^a =$$
$$\begin{bmatrix} \cos\beta_a \cos\psi_a + \sin\beta_a \sin\alpha_a \sin\psi_a & \cos\alpha_a \sin\psi_a & \sin\beta_a \cos\psi_a - \cos\beta_a \sin\alpha_a \sin\psi_a \\ -\cos\beta_a \sin\psi_a + \sin\beta_a \sin\alpha_a \cos\psi_a & \cos\alpha_a \cos\psi_a & -\sin\beta_a \sin\psi_a - \cos\beta_a \sin\alpha_a \cos\psi_a \\ -\sin\beta_a \cos\alpha_a & \sin\alpha_a & \cos\beta_a \cos\alpha_a \end{bmatrix} \quad (11)$$

where $\psi_a$, $\alpha_a$ and $\beta_a$ are yaw, pitch and roll angles in solar azimuth coordinate, respectively. In practice, ENU coordinates are also often used, and the rotation matrix from the imager-centered coordinate system to the ENU coordinate system is

$$C_c^g =$$
$$\begin{bmatrix} \cos\beta_g \cos\psi_g + \sin\beta_g \sin\alpha_g \sin\psi_g & \cos\alpha_g \sin\psi_g & \sin\beta_g \cos\psi_g - \cos\beta_g \sin\alpha_g \sin\psi_g \\ -\cos\beta_g \sin\psi_g + \sin\beta_g \sin\alpha_g \cos\psi_g & \cos\alpha_g \cos\psi_g & -\sin\beta_g \sin\psi_g - \cos\beta_g \sin\alpha_g \cos\psi_g \\ -\sin\beta_g \cos\alpha_g & \sin\alpha_g & \cos\beta_g \cos\alpha_g \end{bmatrix} \quad (12)$$

where $\psi_g$, $\alpha_g$ and $\beta_g$ are yaw, pitch and roll angles in solar azimuth coordinate, respectively.

And the relation between $C_c^a$ and $C_c^g$ is

$$C_c^g = \begin{bmatrix} \cos\varphi_{gS} & \sin\varphi_{gS} & 0 \\ -\sin\varphi_{gS} & \cos\varphi_{gS} & 0 \\ 0 & 0 & 1 \end{bmatrix} C_c^a \quad (13)$$

Next, we further describe the polarization imaging simulation system in the solar azimuth coordinate system. The simulation system in the ENU coordinate system can be obtained by coordinate conversion according to equation (10).

$\overrightarrow{o_c P_c'}$ in solar azimuth coordinate system is

$$\overrightarrow{o_c P_a'} = C_c^a \overrightarrow{o_c P_c'} \quad (14)$$

So, the azimuth angle $\varphi_{aP'}$ of $\overrightarrow{o_c P_a'}$ is

$$\varphi_{aP'} = \arctan\left( \frac{\overrightarrow{o_c P_a'}(1,1)}{\overrightarrow{o_c P_a'}(2,1)} \right) \quad (15)$$

The zenith angle $\theta_{P'}$ of $\overrightarrow{o_c P_a'}$ is

$$\theta_{P'} = \arccos\left( \frac{\overrightarrow{o_c P_a'}(3,1)}{\left| \overrightarrow{o_c P_a'} \right|} \right) \quad (16)$$

where $\overrightarrow{o_c P_a'}(1,1)$, $\overrightarrow{o_c P_a'}(2,1)$ and $\overrightarrow{o_c P_a'}(3,1)$ represent the components of $\overrightarrow{o_c P_a'}$, $\left| \overrightarrow{o_c P_a'} \right|$ denotes the mode of $\overrightarrow{o_c P_a'}$. Then, the scattering angle $\gamma_{P'}$ of $P'$ is given by

$$\gamma_{P'} = \arccos\left( \cos\theta_{P'} \cos\theta_S + \sin\theta_{P'} \sin\theta_S \cos(\varphi_{aP'}) \right) \quad (17)$$

Combining equation (1), (16) and (17), the total $LI_{P'}$ of $P'$ is given.

The projection of $P'$ on complex coordinate system is

$$\begin{cases} \mathrm{Re}[\zeta_{P'}] = \tan\left( \frac{1}{2} \theta_{P'} \right) \cos(\varphi_{aP'}) \\ \mathrm{Im}[\zeta_{P'}] = \tan\left( \frac{1}{2} \theta_{P'} \right) \sin(\varphi_{aP'}) \end{cases} \quad (18)$$

where $\zeta_{P'}$ is the projection of $P'$ on the complex coordinate system.

Combining equation (5), (16), (17) and (18), the $DOP_{P'}$ of $P'$ is given.

The value of AOP is determined by the reference direction, so the polarization E-vector corresponding to $P'$ is first obtained. Combining equation (3), (9), (16), (17) and (18), the polarization E-vector $\overrightarrow{E_{aP'}}$ of $P'$ in solar azimuth coordinate system is given. Then, the polarization E-vector $\overrightarrow{E_{cP'}}$ of $P'$ in imager-centered coordinate system is given by

$$\overrightarrow{E_{cP'}} = C_a^c \overrightarrow{E_{aP'}} \quad (19)$$

The AOP reference direction of the imager coincides with $y_b$ axis and the optical axis of the imager coincides with $z_c$ axis. AOP is the angle between polarization E-vector and reference direction. So, AOP of $P'$ captured by polarization imager is given by

$$AOP_c = \arctan\frac{\overrightarrow{E_{cP'}}(1,1)}{\overrightarrow{E_{cP'}}(2,1)} \quad (20)$$

where $\overrightarrow{E_{cP'}}(1,1)$, $\overrightarrow{E_{cP'}}(2,1)$ represents the components of $\overrightarrow{E_{cP'}}$,

and the reference direction of $AOP_c$ is $y_b$ axis.

Above all, hypothetical LI, DOP and $AOP_c$ images are captured.

### B. Original light intensity imager

However, the DOP and $AOP_c$ images are not directly captured by polarization imager, which are obtained by fusing several original black-and-white LI images in different polarization directions. To construct a complete polarization imaging simulation system, the original LI images in different polarization directions are captured in this section.

The Stokes vector $(LI, Q, U, V)$ is usually defined to describe the polarized light, where $LI$ describes the total light intensity, $Q$ describes the preponderance of linearly horizontal polarized light over linearly vertical polarized light, $U$ describes the preponderance of linearly +45° polarized light over linearly -45° polarized light, $V$ describes the preponderance of right-handed polarized light over left-handed polarized light. The scattered skylight is seen to be linearly polarized in the atmosphere [31, 34], so $V = 0$. Then, according to the definition of Stokes vector, we have

$$\begin{cases} LI = \dfrac{1}{2}\big[LI(0°) + LI(45°) + LI(90°) + LI(135°)\big] \\ Q = LI(0°) - LI(90°) \\ U = LI(45°) - LI(135°) \\ V = 0 \end{cases} \quad (21)$$

where $LI(0°)$, $LI(45°)$, $LI(90°)$, $LI(135°)$ are the LI in the $0°$, $45°$, $90°$, and $135°$ polarization directions, respectively. So, the DOP and $AOP_c$ captured by polarization imager can be expressed by

$$\begin{cases} DOP = \dfrac{\sqrt{Q^2 + U^2}}{LI} \\ AOP_c = \dfrac{1}{2}\arctan\left(\dfrac{U}{Q}\right) \end{cases} \quad (22)$$

Rewrite equation (22), $Q$ and $U$ are given by

$$\begin{cases} Q = DOP \cdot LI \cdot \cos(2AOP_c) \\ U = DOP \cdot LI \cdot \sin(2AOP_c) \end{cases} \quad (23)$$

Substituting (23) into (21), the original LI in different directions are given by

$$\begin{pmatrix} LI(0°) \\ LI(45°) \\ LI(90°) \\ LI(135°) \end{pmatrix} = \dfrac{\kappa}{2}\begin{pmatrix} 1 & 1 & 0 \\ 1 & 0 & 1 \\ 1 & -1 & 0 \\ 1 & 0 & -1 \end{pmatrix}\begin{pmatrix} LI \\ Q \\ U \end{pmatrix}$$

$$= \dfrac{\kappa}{2}\begin{pmatrix} 1 & 1 & 0 \\ 1 & 0 & 1 \\ 1 & -1 & 0 \\ 1 & 0 & -1 \end{pmatrix}\begin{pmatrix} LI \\ DOP \cdot LI \cdot \cos(2AOP_c) \\ DOP \cdot LI \cdot \sin(2AOP_c) \end{pmatrix} \quad (24)$$

where $\kappa$ is defined to express the total influence of imager parameters, such as sensor gain, exposure time, aperture size and so on. In practice, the gray value of original black-and-white LI images is generally between 0 and 255, so the value of k should be guaranteed the value of $LI(0°)$, $LI(45°)$, $LI(90°)$, $LI(135°)$ between 0 and 255.

In short, substituting LI, DOP and $AOP_c$ into equation (24), the original black-and-white LI images in $0°$, $45°$, $90°$, and $135°$ polarization directions are captured.

## IV. SIMULATION DATASET

### A. Simulation results of skylight polarization pattern

Fig. 5, 6, 7, 8, and 9 show several samples of the designed skylight polarization model described in section 2, for systematic variations in solar azimuth angle, solar zenith angle, atmospheric turbidity, wavelength and ground albedo, respectively. In these figures, except for the varying parameter, the other parameters remain fixed as follows: solar azimuth angle is 200°, solar zenith angle is 50°, atmospheric turbidity is 4, wavelength is 450nm, ground albedo is 0.1. The white dotted line in Fig. 5, 6, 7, 8, and 9 represents SM-ASM.

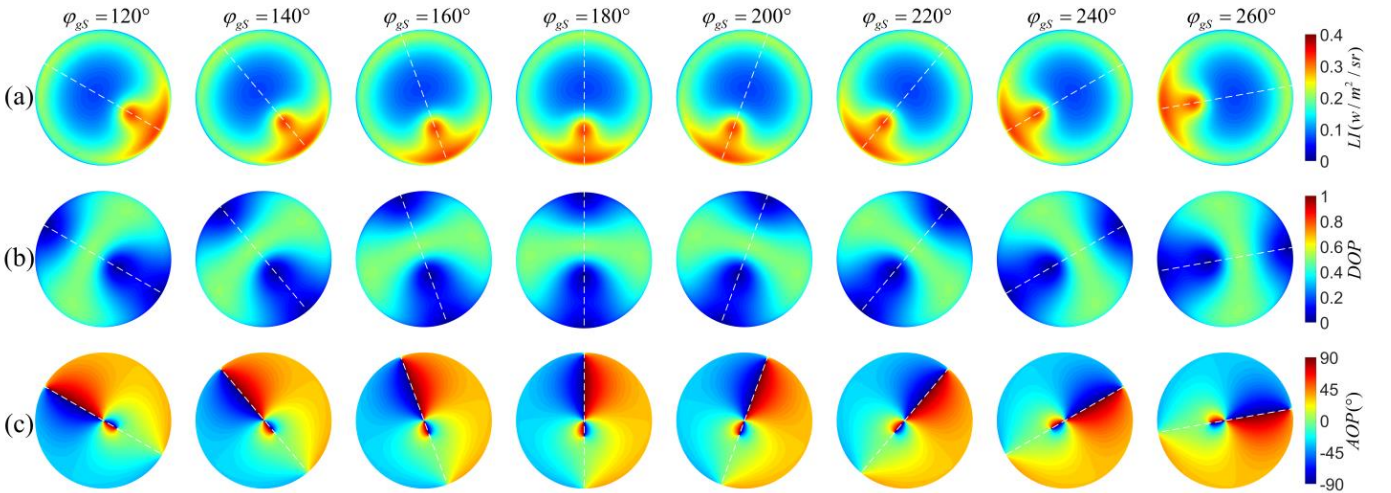

Fig. 5. Simulation results of skylight polarization pattern are varied for the solar azimuth angle $\varphi_{gs}$ in ENU coordinate system: (a) LI of skylight polarization pattern; (b) DOP of skylight polarization pattern; (c) AOP of skylight polarization pattern. Solar zenith angle is 50°, atmospheric turbidity is 4, wavelength is 450nm, ground albedo is 0.1. The white dotted line represents SM-ASM.

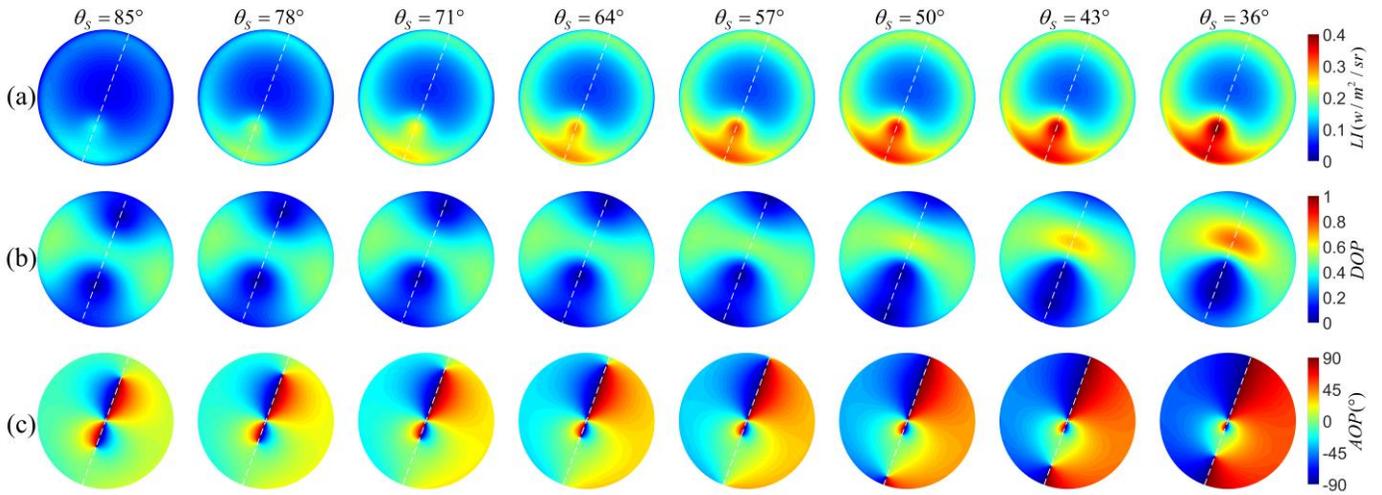

Fig. 6. Simulation results of skylight polarization pattern are varied for the solar zenith angle $\theta_S$ : (a) LI of skylight polarization pattern; (b) DOP of skylight polarization pattern; (c) AOP of skylight polarization pattern. Solar azimuth angle is 200°, atmospheric turbidity is 4, wavelength is 450nm, ground albedo is 0.1. The white dotted line represents SM-ASM.

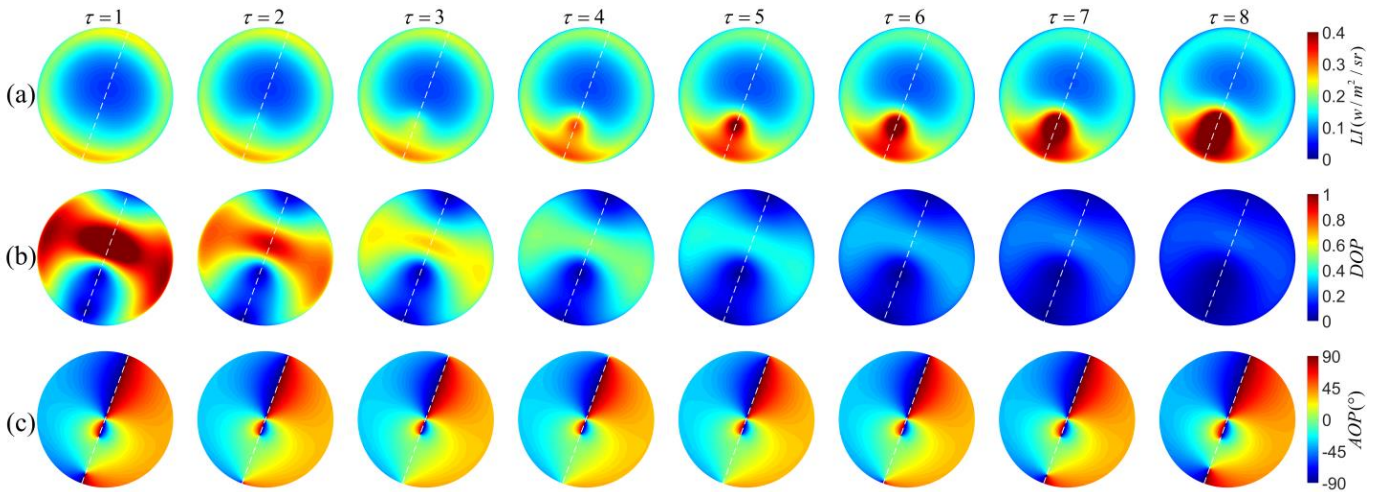

Fig. 7. Simulation results of skylight polarization pattern are varied for the atmospheric turbidity $\tau$ : (a) LI of skylight polarization pattern; (b) DOP of skylight polarization pattern; (c) AOP of skylight polarization pattern. Solar azimuth angle is 200°, solar zenith angle is 50°, wavelength is 450nm, ground albedo is 0.1. The white dotted line represents SM-ASM.

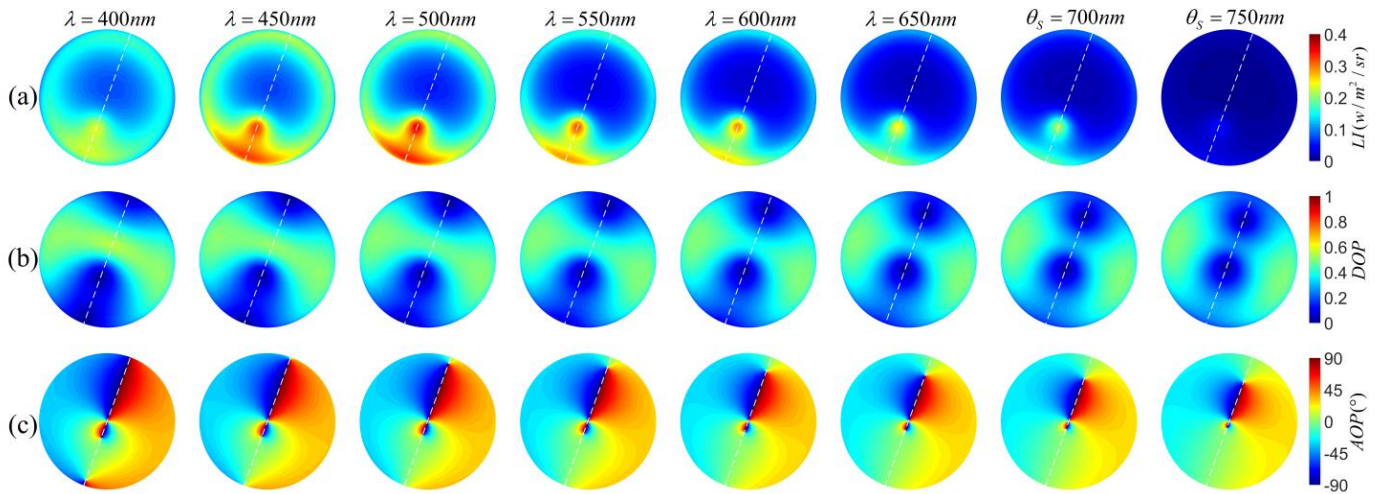

Fig. 8. Simulation results of skylight polarization pattern are varied for the wavelength $\lambda$ : (a) LI of skylight polarization pattern; (b) DOP of skylight polarization pattern; (c) AOP of skylight polarization pattern. Solar azimuth angle is 200°, solar zenith angle is 50°, atmospheric turbidity is 4, ground albedo is 0.1. The white dotted line represents SM-ASM.

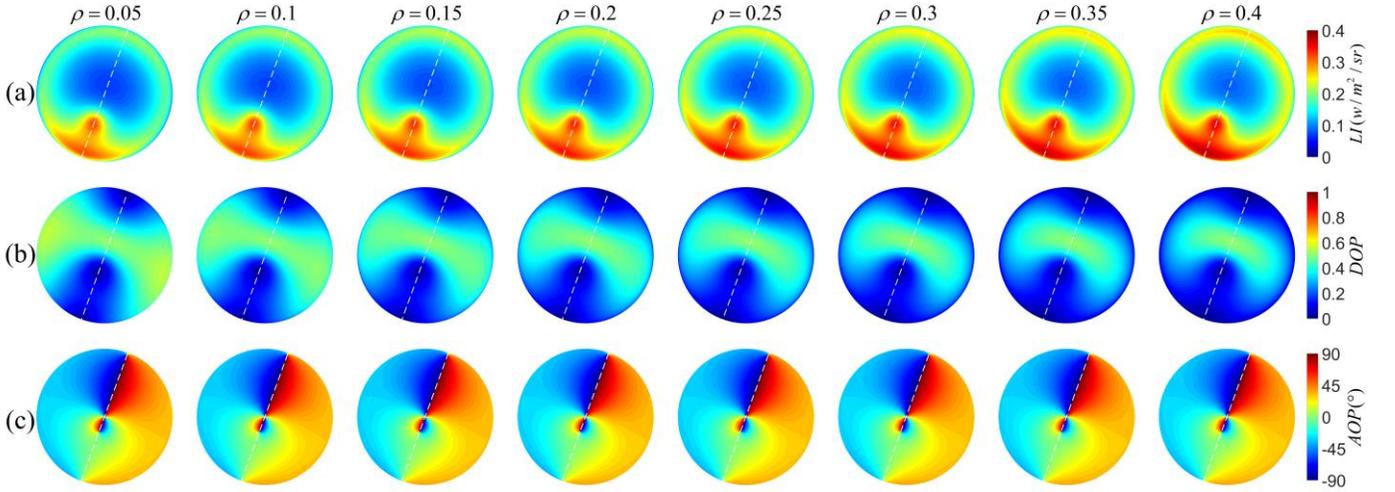

Fig. 9. Simulation results of skylight polarization pattern are varied for the ground albedo $\rho$: (a) LI of skylight polarization pattern; (b) DOP of skylight polarization pattern; (c) AOP of skylight polarization pattern. Solar azimuth angle is 200°, solar zenith angle is 50°, atmospheric turbidity is 4, wavelength is 450nm. The white dotted line represents SM-ASM.

As illustrated in Fig. 5, the designed model well describes the symmetry of skylight polarization pattern. Skylight polarization pattern is symmetric about SM-ASM and rotates with the change of solar azimuth angle [22].

The existence of neutral points can be seen on DOP and AOP of skylight polarization model. Arago and Babinet neural points can be observed as the Sun closed to the horizon, while Brewster and Babinet2 are below the horizon, as illustrated in Fig. 6 under $\theta_s$ =85°, 78° and 71°. As the sun rise, Arago neural point disappears and Brewster neural point appears, Brewster and Babinet neural points are above the horizon, while Arago and Babinet2 are below the horizon, illustrated in Fig. 6 under $\theta_s$ =50°, 43° and 36°. In addition, as shown in Fig. 6(c) under $\theta_s$ =85°, 78°, the AOP of skylight polarization model presents an obvious 8-shape, which is the same as the observation [62].

With the increase of atmospheric turbidity, the influence of Mie scattering and multi-scattering increases, resulting in an increase in LI and a decrease in DOP of skylight polarization pattern [43]. As illustrated in Fig. 7, for the ideal case of $\tau$ =1, the maximum DOP of skylight polarization pattern is 1. However, the DOP of the whole sky decreases sharply with the increase of atmospheric turbidity.

The skylight polarization patterns of different wavelengths have some differences, as illustrated in Fig. 8. In addition, the LI of blue light is stronger than that of other wavelengths, which is consistent with the fact that the sky is blue [63].

Sunlight reflected from the Earth's surface increases the LI and weakens the polarization properties of skylight near the Earth's surface [44]. Increasing the value of ground albedo results in an increase in LI and a decrease in DOP near the horizon, as illustrated in Fig. 9.

### B. Simulation results of polarization imaging system

As described above, not only DOP and AOP$_c$ images, but also the original LI images in different polarization directions can be captured by the designed polarization imaging system. The parameters of polarization imager are shown in Table I.

TABLE I
PARAMETERS OF POLARIZATION IMAGER

| Symbol | Value | Units | Description |
| --- | --- | --- | --- |
| $f$ | 4 | mm | Focal length of polarization imager |
| $\lambda_c$ | 380-780 | nm | Wavelength range of polarization imager |
| $\eta_x$ | 2048 | pixel | Number of pixels in $x_p$ direction |
| $\eta_y$ | 2448 | pixel | Number of pixels in $y_p$ direction |
| $D_x$ | 3.45 | μm | Pixel size in $x_p$ direction |
| $D_y$ | 3.45 | μm | Pixel size in $y_p$ direction |

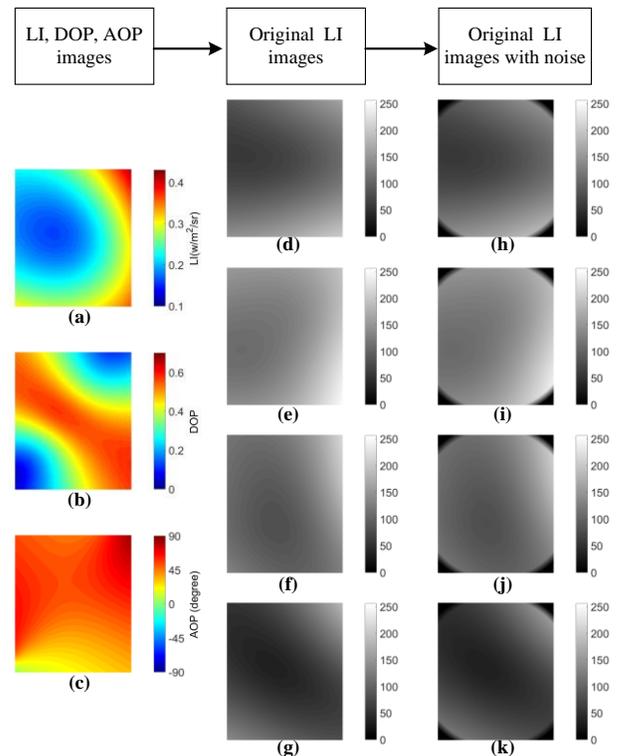

Fig. 10. (a) LI image; (b) DOP image; (c) AOP$_c$ image; (d) 0° original LI image; (e) 45° original LI image; (f) 90° original LI image; (g) 135° original LI image; (h) 0° original LI image with noise; (i) 45° original LI image with noise; (j) 90° original LI image with noise; (k) 135° original LI image with noise.

As illustrated in Fig. 10, Fig. 10(d), Fig. 10(e), Fig. 10(f) and Fig. 10(g) represent original LI images in the 0°, 45°, 90°, and 135° polarization directions respectively, which are captured by equation (24). The main advantage of capturing original LI images is that it can show the complete process of the polarization imaging. When capturing skylight polarization pattern, the original LI images in different polarization directions are first captured, and then Stokes vector is calculated to capture DOP and $AOP_c$ images. In addition, the noises caused by measurement, insufficient illumination at the edge of imager field of view [64] and influence of clouds are added on original LI images, then the original LI images with noise are captured as shown in Fig. 10(h), Fig. 10(i), Fig. 10(j) and Fig. 10(k), for example, Gaussian white noise and insufficient illumination at the edge of imager field of view are added.

Above all, as shown in Fig. 1, to construct PSNS dataset, influence of sun position, neutral points, atmospheric turbidity, wavelength and ground albedo are considered to capture original LI images in different polarization directions. Based on

PHX050S-P Sony IMX250MZR COMS polarization imager, an initial version of the PSNS dataset has been constructed. The initial version of the dataset can be downloaded here:

https://pan.baidu.com/s/1Ck1Kpojh0tON4S1C5wu4pA
Password: psns

There are about 138000 images in this dataset, and this dataset is divided into three parts, which are placed in different folders: noise free polarization images (NFPI); Gaussian noise and insufficient illumination polarization images (GNIIPI); Gaussian noise, insufficient illumination, and clouds polarization images (GNIICPI). The polarization images in NFPI folder have no noise and are ideal polarization images. The polarization images in GNIIPI folder have 2% Gaussian white noise and insufficient illumination polarization at the edge of images. The polarization images in GNIICPI folder have 2% Gaussian white noise, insufficient illumination polarization at the edge of images, and influence of clouds. The cloud interference is simulated by the superposition of multiple ellipses. In the area where the ellipse is located, the gray value of pixel is set to 255. And in order to facilitate testing, test sets have been added: Test set of noise free polarization images (TNFPI), as shown in Fig. 11; Test set of Gaussian noise and insufficient illumination polarization images (TGNIIPI), as shown in Fig. 12; Test set of Gaussian noise, insufficient illumination, and clouds polarization images (TGNIICPI), as shown in Fig. 13. And we will continue to update and expand PSNS dataset.

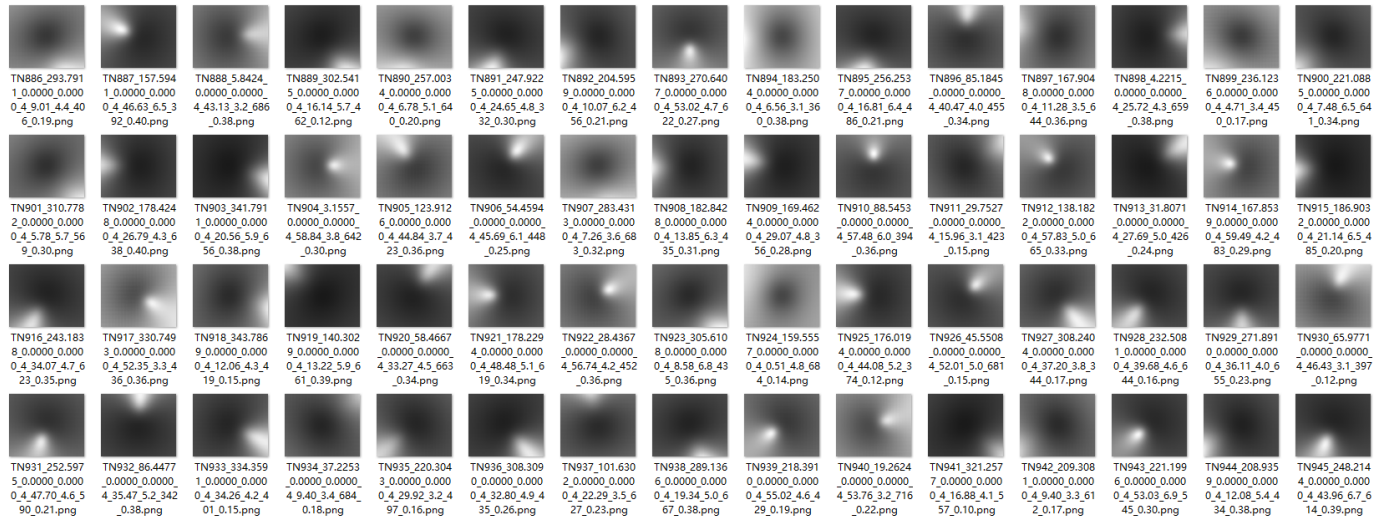

Fig. 11. Polarization images in TNFPI (Test set of noise free polarization images) folder of PSNS dataset.

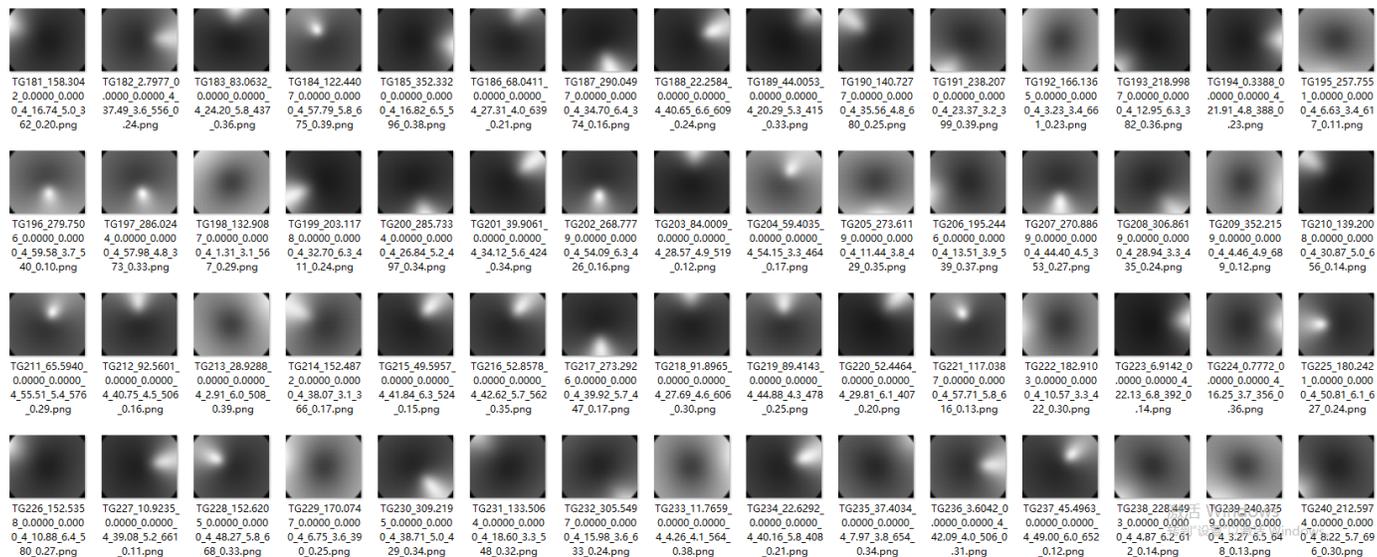

Fig. 12. Polarization images in TGNIIPI (Test set of Gaussian noise and insufficient illumination polarization images) folder of PSNS dataset.

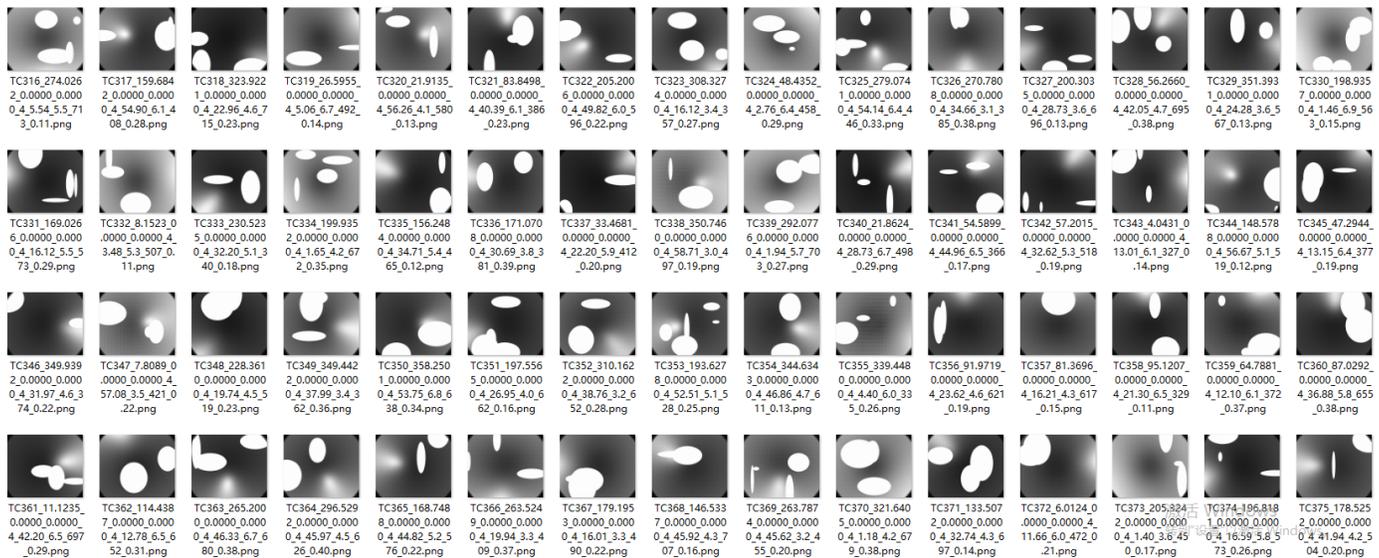

Fig. 13. Polarization images in TGNIICPI (Test set of Gaussian noise, insufficient illumination, and clouds polarization images) folder of PSNS dataset.

In addition, because of a variety of polarization sensors and sky models, to facilitate researchers to build their own datasets based on their own polarization sensors and sky models, we have disclosed the source code of polarization imager and original LI imager on GitHub. The code of polarization imager refers to our previous work [48].

https://github.com/HuajuLiang/HypotheticalPolarizationCamera

The code of original light intensity imager is available at:

https://github.com/HuajuLiang/OriginalLightItensity

## V. CONCLUSION

In this paper, an open PSNS dataset is constructed for the first time based on skylight polarization model and polarization imaging system. In addition, the major contribution of this paper is that we have uploaded polarization imager code and original light intensity imager code to GitHub, to facilitate researchers to build their own datasets based on their own polarization sensors and models. All these will greatly promote the application of machine learning methods in the field of polarized skylight navigation.

Although, the simulation original black-and-white LI images in different directions are captured based on the designed sky model and polarization imaging simulation system to construct PSNS dataset, the actual dataset still needs to be built. Perfect actual datasets need to consider different weather, different time, different geographical locations and so on. It is time-consuming and laborious, which requires the efforts of all researchers in this field.

## REFERENCES

[1] Z. Li *et al.*, "Remote sensing of atmospheric particulate mass of dry PM2.5 near the ground: Method validation using ground-based measurements," *Remote Sensing of Environment,* vol. 173, pp. 59-68, 2016, doi: 10.1016/j.rse.2015.11.019.

[2] J.-P. Gastellu-Etchegorry *et al.*, "DART: Recent Advances in Remote Sensing Data Modeling With Atmosphere, Polarization, and Chlorophyll Fluorescence," *IEEE Journal of Selected Topics in Applied Earth Observations and Remote Sensing,* vol. 10, no. 6, pp. 2640-2649, 2017, doi: 10.1109/jstars.2017.2685528.

[3] W. Chen *et al.*, "Aerosol-induced changes in sky polarization pattern: potential hint on applications in polarimetric remote sensing," *International Journal of Remote Sensing,* pp. 1-18, 2019, doi: 10.1080/01431161.2019.1685724.

[4] P. Du, A. Samat, B. Waske, S. Liu, and Z. Li, "Random Forest and Rotation Forest for fully polarized SAR image classification using polarimetric and spatial features," *ISPRS Journal of Photogrammetry and Remote Sensing,* vol. 105, pp. 38-53, 2015, doi: 10.1016/j.isprsjprs.2015.03.002.

[5] M. Ahishali, S. Kiranyaz, T. Ince, and M. Gabbouj, "Dual and Single Polarized SAR Image Classification Using Compact Convolutional Neural Networks," *Remote Sensing,* vol. 11, no. 11, 2019, doi: 10.3390/rs11111340.

[6] J. Ayers, E. Gkanias, B. Risse, M. Mangan, and B. Webb, "From skylight input to behavioural output: A computational model of the insect polarised light compass," *PLOS Computational Biology,* vol. 15, no. 7, 2019, doi: 10.1371/journal.pcbi.1007123.

[7] J. Dupeyroux, J. R. Serres, and S. Viollet, "AntBot: A six-legged walking robot able to home like desert ants in outdoor environments," *Science Robotics,* vol. 4, no. 0307, p. 12, 2019, doi: 10.1126/scirobotics.aau0307

[8] Z. Wan, K. Zhao, and J. Chu, "Robust azimuth measurement method based on polarimetric imaging for bionic polarization navigation," *IEEE Transactions on Instrumentation and Measurement,* pp. 1-1, 2020, doi: 10.1109/tim.2019.2959291.

[9] M. Müller and R. Wehner, "Path integration provides a scaffold for landmark learning in desert ants," *Curr Biol,* vol. 20, no. 15, pp. 1368-71, Aug 10 2010, doi: 10.1016/j.cub.2010.06.035.

[10] F. Lebhardt and B. Ronacher, "Interactions of the polarization and the sun compass in path integration of desert ants," *J Comp Physiol A Neuroethol Sens Neural Behav Physiol,* vol. 200, no. 8, pp. 711-20, Aug 2014, doi: 10.1007/s00359-013-0871-1.

[11] M. Dacke, B. el Jundi, J. Smolka, M. Byrne, and E. Baird, "The role of the sun in the celestial integration of dung beetles," *Philos Trans R Soc Lond B Biol Sci,* vol. 369, no. 1636, p. 20130036, 2014, doi: 10.1098/rstb.2013.0036.

[12] J. Smolka, E. Baird, B. el Jundi, T. Reber, M. J. Byrne, and M. Dacke, "Night sky orientation with diurnal and nocturnal eyes: dim-light

adaptations are critical when the moon is out of sight," *Animal Behaviour,* vol. 111, pp. 127-146, 2016, doi: 10.1016/j.anbehav.2015.10.005.

[13] T. L. Warren, P. T. Weir, and M. H. Dickinson, "Flying Drosophilamelanogaster maintain arbitrary but stable headings relative to the angle of polarized light," *J Exp Biol,* vol. 221, no. Pt 9, May 11 2018, doi: 10.1242/jeb.177550.

[14] T. L. Warren, Y. M. Giraldo, and M. H. Dickinson, "Celestial navigation in Drosophila," *J Exp Biol,* vol. 222, no. Pt Suppl 1, Feb 6 2019, doi: 10.1242/jeb.186148.

[15] W. Zhi, J. Chu, J. Li, and Y. Wang, "A Novel Attitude Determination System Aided by Polarization Sensor," *Sensors (Basel),* vol. 18, no. 1, Jan 9 2018, doi: 10.3390/s18010158.

[16] M. Hamaoui, "Polarized skylight navigation," *Appl Opt,* vol. 56, no. 3, pp. B37-B46, Jan 20 2017, doi: 10.1364/AO.56.000B37.

[17] B. P. Samuel, G. Roman, M. Justin, R. Charbel, and G. Viktor, "Bioinspired polarization vision enables underwater geolocalization," *Science Advances,* vol. 4, 2018, doi: 10.1126/sciadv.aao6841.

[18] G. Zhou, J. Wang, W. Xu, K. Zhang, and Z. Ma, "Polarization Patterns of Transmitted Celestial Light under Wavy Water Surfaces," *Remote Sensing,* vol. 9, no. 4, 2017, doi: 10.3390/rs9040324.

[19] J. Dupeyroux, S. Viollet, and J. R. Serres, "An ant-inspired celestial compass applied to autonomous outdoor robot navigation," *Robotics and Autonomous Systems,* vol. 117, pp. 40-56, 2019, doi: 10.1016/j.robot.2019.04.007.

[20] K. D. Pham, G. Chen, T. Aycock, A. Lompado, T. Wolz, and D. Chenault, "Passive optical sensing of atmospheric polarization for GPS denied operations," presented at the Sensors and Systems for Space Applications IX, 2016, 2016.

[21] Z. Huijie *et al.,* "Polarization patterns under different sky conditions and a navigation method based on the symmetry of the AOP map of skylight," *Optics Express,* vol. 26, no. 22, 2018, doi: 10.1364/OE.26.028589.

[22] W. Zhang, Y. Cao, X. Zhang, Y. Yang, and Y. Ning, "Angle of sky light polarization derived from digital images of the sky under various conditions," *Appl Opt,* vol. 56, no. 3, pp. 587-595, Jan 20 2017, doi: 10.1364/AO.56.000587.

[23] Y. Wang, L. Zhang, X. He, X. Hu, and C. Fan, "Multicamera polarized vision for the orientation with the skylight polarization patterns," *Optical Engineering,* vol. 57, no. 04, 2018, doi: 10.1117/1.Oe.57.4.043101.

[24] Y. Wang, J. Chu, R. Zhang, and C. Shi, "Orthogonal vector algorithm to obtain the solar vector using the single-scattering Rayleigh model," *Appl Opt,* vol. 57, no. 4, pp. 594-601, Feb 1 2018, doi: 10.1364/AO.57.000594.

[25] J. Yang, T. Du, X. Liu, B. Niu, and L. Guo, "Method and Implementation of a Bioinspired Polarization-Based Attitude and Heading Reference System by Integration of Polarization Compass and Inertial Sensors," *IEEE Transactions on Industrial Electronics,* vol. 67, no. 11, pp. 9802-9812, 2020, doi: 10.1109/tie.2019.2952799.

[26] L. Guan *et al.,* "Study on skylight polarization patterns over the ocean for polarized light navigation application," *Appl Opt,* vol. 57, no. 21, pp. 6243-6251, Jul 20 2018, doi: 10.1364/AO.57.006243.

[27] H. Lu, K. Zhao, Z. You, and K. Huang, "Angle algorithm based on Hough transform for imaging polarization navigation sensor," *Opt Express,* vol. 23, no. 6, pp. 7248-62, Mar 23 2015, doi: 10.1364/OE.23.007248.

[28] X. Wang, J. Gao, and N. W. Roberts, "Bio-inspired orientation using the polarization pattern in the sky based on artificial neural networks," *Opt Express,* vol. 27, no. 10, pp. 13681-13693, May 13 2019, doi: 10.1364/OE.27.013681.

[29] E. Gkanias, B. Risse, M. Mangan, and B. Webb, "From skylight input to behavioural output: A computational model of the insect polarised light compass," *PLoS Comput Biol,* vol. 15, no. 7, p. e1007123, Jul 2019, doi: 10.1371/journal.pcbi.1007123.

[30] A. K. Alexander, *AEROSOL OPTICS Light Absorption and Scattering by Particles in the Atmosphere.* Germany, Berlin: Springer, 2008.

[31] A. K. Alexander, *Light Scattering Reviews 10.* Berlin, Germany: Springer, 2016.

[32] M. Carlos, O. Takeshi, and I. Katsushi, "Real-time rendering of aerial perspective effect based on turbidity estimation," *IPSJ Transactions on Computer Vision and Applications,* vol. 9, no. 1, 2017, doi: 10.1186/s41074-016-0012-1.

[33] J. Dupeyroux, S. Viollet, and J. R. Serres, "Polarized skylight-based heading measurements: a bio-inspired approach," *J R Soc Interface,* vol. 16, no. 150, p. 20180878, Jan 31 2019, doi: 10.1098/rsif.2018.0878.

[34] C. F. Bohren, *Atmospheric Optics.* Pennsylvania Pennsylvania State University, 2007.

[35] Z. Kaichun, C. Jinkui, W. Tichang, and Z. Qiang, "A Novel Angle Algorithm of Polarization Sensor for Navigation," *IEEE Transactions on Instrumentation and Measurement,* vol. 58, no. 8, pp. 2791-2796, 2009, doi: 10.1109/tim.2009.2016299.

[36] I. M. Michael, D. T. Larry, and A. L. Andrew, *Scattering, Absorption, and Emission of Light by Small Particles.* Cambridge, UK: Cambridge University Press, 2002.

[37] C. Davis, C. Emde, and R. Harwood, "A 3-D polarized reversed Monte Carlo radiative transfer model for Millimeter and submillimeter passive remote sensing in cloudy atmospheres," *IEEE Transactions on Geoscience and Remote Sensing,* vol. 43, no. 5, pp. 1096-1101, 2005, doi: 10.1109/tgrs.2004.837505.

[38] C. Xiong, J. Shi, D. Ji, T. Wang, Y. Xu, and T. Zhao, "A New Hybrid Snow Light Scattering Model Based on Geometric Optics Theory and Vector Radiative Transfer Theory," *IEEE Transactions on Geoscience and Remote Sensing,* vol. 53, no. 9, pp. 4862-4875, 2015, doi: 10.1109/tgrs.2015.2411592.

[39] G. Horváth, B. Balázs, S. Bence, and B. András, "First observation of the fourth neutral polarization point in the atmosphere," *Optical Society of America,* vol. 19, no. 10, 2002.

[40] M. V. Berry, M. R. Dennis, and R. L. Lee Jr, "Polarization singularities in the clear sky," *New Journal of Physics,* vol. 6, no. 162, 2004, doi: 10.1088/1367-2630/6/1/162.

[41] M. V. Berry, "Nature's optics and our understanding of light," *Contemporary Physics,* vol. 56, pp. 1-15, 2014, doi: 10.1080/00107514.2015.971625.

[42] J. H. Hannay, "Polarization of sky light from a canopy atmosphere," *New Journal of Physics,* vol. 6, no. 197, 2004, doi: 10.1088/1367-2630/6/1/197.

[43] A. Wilkie, F. T. Robert, C. Ulbricht, G. Zotti, and W. Purgathofer, "An Analytical Model for Skylight Polarisation," presented at the Eurographics Symposium on Rendering, 2004.

[44] X. Wang, J. Gao, Z. Fan, and N. W. Roberts, "An analytical model for the celestial distribution of polarized light, accounting for polarization singularities, wavelength and atmospheric turbidity," *Journal of Optics,* vol. 18, no. 6, 2016, doi: 10.1088/2040-8978/18/6/065601.

[45] H. Lukas and W. Alexander, "An analytic model for full spectral sky-dome radiance," *ACM Transactions on Graphics (TOG) - SIGGRAPH 2012 Conference Proceedings,* 2012, doi: 10.1145/2185520.2185591.

[46] H. Lukas and W. Alexander, "Adding a Solar-Radiance Function to the Hošek-Wilkie Skylight Model," *IEEE engineering in medicine and biology magazine: the quarterly magazine of the Engineering in Medicine & Biology Society,* vol. 33, no. 3, pp. 44-55, 2013.

[47] J. T. Kider, D. Knowlton, J. Newlin, Y. K. Li, and D. P. Greenberg, "A framework for the experimental comparison of solar and skydome illumination," *ACM Transactions on Graphics,* vol. 33, no. 6, pp. 1-12, 2014, doi: 10.1145/2661229.2661259.

[48] H. Liang, H. Bai, N. Liu, and K. Shen, "Limitation of Rayleigh sky model for bio-inspired polarized skylight navigation in three-dimensional attitude determination," *Bioinspir Biomim,* Feb 27 2020, doi: 10.1088/1748-3190/ab7ab7.

[49] C.-Y. Liu and L.-J. Chang, "Characterization of Surface Micro-Roughness by Off-Specular Measurements of Polarized Optical Scattering," *Measurement Science Review,* vol. 19, no. 6, pp. 257-263, 2019, doi: 10.2478/msr-2019-0033.

[50] A. Dashpute, C. Anand, and M. Sarkar, "Depth Resolution Enhancement in Time-of-Flight Cameras Using Polarization State of the Reflected Light," *IEEE Transactions on Instrumentation and Measurement,* vol. 68, no. 1, pp. 160-168, 2019, doi: 10.1109/tim.2018.2838819.

[51] Q. Tan, Q. Xu, N. Xie, and C. Li, "A New Optical Voltage Sensor Based on Radial Polarization Detection," *IEEE Transactions on Instrumentation and Measurement,* vol. 66, no. 1, pp. 158-164, 2017, doi: 10.1109/tim.2016.2621198.

[52] J. Ko, "A very simple and accurate way to measure the transmission axis of a linear polarizer," *Measurement Science and Technology,* vol. 31, no. 1, 2020, doi: 10.1088/1361-6501/ab3688.

[53] T. Deng, J. Zeng, S. Wang, S. Yan, and A. Chen, "An optical fire detector with enhanced response sensitivities for black smoke based on the polarized light scattering," *Measurement Science and Technology,* vol. 30, no. 11, 2019, doi: 10.1088/1361-6501/ab2e35.

[54] C. Fan, X. Hu, X. He, L. Zhang, and J. Lian, "Integrated Polarized Skylight Sensor and MIMU With a Metric Map for Urban Ground Navigation," *IEEE Sensors Journal,* vol. 18, no. 4, pp. 1714-1722, 2018, doi: 10.1109/jsen.2017.2786404.

[55] M. Sarkar, D. San Segundo Bello, C. Van Hoof, and A. J. P. Theuwissen, "Integrated Polarization-Analyzing CMOS Image Sensor for Detecting the Incoming Light Ray Direction," *IEEE Transactions on*

*Instrumentation and Measurement,* vol. 60, no. 8, pp. 2759-2767, 2011, doi: 10.1109/tim.2011.2130050.

[56] Z. Xiaojin, L. Xin, A. Abubakar, and B. Amine, "Novel micro-polarizer array patterns for CMOS polarization image sensors," presented at the International Conference on Electronic Devices, Ras Al Khaimah, United Arab Emirates, 2017.

[57] A. Ashfaq, Z. Xiaojin, and B. Amine, "Polarization Imaging for Remote Sensing " presented at the IEEE Microwaves, Radar and Remote Sensing Symposium (MRRS), 2017.

[58] B. Bernard, "Improvement in solar declination computation," *Solar Energy,* vol. 35, no. 4, pp. 367-369, 1985.

[59] I. Reda and A. Andreas, "Solar position algorithm for solar radiation applications," *Solar Energy,* vol. 76, no. 5, pp. 577-589, 2004, doi: 10.1016/j.solener.2003.12.003.

[60] G. Roberto, "An algorithm for the computation of the solar position," *Solar Energy,* vol. 82, pp. 462-470, 2008, doi: 10.1016/j.solener.2007.10.001.

[61] R. Grena, "Five new algorithms for the computation of sun position from 2010 to 2110," *Solar Energy,* vol. 86, no. 5, pp. 1323-1337, 2012, doi: 10.1016/j.solener.2012.01.024.

[62] J. Gál, G. Horváth, A. Barta, and R. Wehner, "Polarization of the moonlit clear night sky measured by full-sky imaging polarimetry at full Moon: Comparison of the polarization of moonlit and sunlit skies," *Journal of Geophysical Research: Atmospheres,* vol. 106, no. D19, pp. 22647-22653, 2001, doi: 10.1029/2000jd000085.

[63] E. O. Hulburt, "Explanation of the Brightness and Color of the Sky, Particularly the Twilight Sky," *JOURNAL OF THE OPTICAL SOCIETY OF AMERICA,* vol. 43, no. 2, 1953.

[64] Y. Daoyin and T. Ying, *Engineering Optics (The Fourth Edition).* China Machine Press, 2016.

Polarized Skylight Navigation Simulation (PSNS) Dataset

*********************************************************************************************
%%%%%%%%   Manuscript   %%%%%%%%%
Polarized Skylight Navigation Simulation (PSNS) Dataset, Huaju Liang and Hongyang Bai, Nanjing University of Science and Technology
The construction process of the dataset can refer to this manuscript.
*********************************************************************************************

*********************************************************************************************
%%%%%%%%   PSNS Dataset   %%%%%%%%%
There are about 138000 images in this dataset, and this dataset is divided into six parts, which are placed in different folders: noise free polarization images (NFPI), Gaussian noise and insufficient illumination polarization images (GNIIPI), gaussian noise, insufficient illumination and clouds polarization images (GNIICPI). test set of noise free polarization images (TNFPI), test set of Gaussian noise and insufficient illumination polarization images (TGNIIPI), test set of gaussian noise, insufficient illumination and clouds polarization images (TGNIICPI).
At present, the data set can be mainly applied to polarized skylight orientation determination without tilt, so the roll and pitch angles are always 0. If the impact of tilts needs to be eliminated or constructing polarized skylight integrated navigation system, additional navigation data detected by other navigation sensors is needed. Therefore, different applications require different data. In addition, because of a variety of polarization sensors and sky models, to facilitate researchers to build their own datasets based on their own polarization sensors and skylight models, we have disclosed the source code of polarization imager and original LI imager on GitHub.
    https://github.com/HuajuLiang/HypotheticalPolarizationCamera
    https://github.com/HuajuLiang/OriginalLightItensity
*********************************************************************************************

*********************************************************************************************
%%%%%%%%   Folder name   %%%%%%%%%%
NFPI, GNIIPI, GNIICPI, TNFPI, TGNIIPI, TGNIICPI.
NFPI represents noise free polarization images. The polarization images in NFPI folder have no noise and are ideal polarization images.
GNIIPI represents Gaussian noise and insufficient illumination polarization images. The polarization images in GNIIPI folder has 2% Gaussian white noise and insufficient illumination polarization at the edge of images.
GNIICPI represents gaussian noise, insufficient illumination and clouds polarization images (GNIICPI). The polarization images in GNIICPI folder have 2% Gaussian white noise, insufficient illumination polarization at the edge of images, and influence of clouds. The cloud interference is simulated by the superposition of multiple ellipses. In the area where the ellipse is located, the gray value of the pixel is set to 255.
TNFPI represents the test set of noise free polarization images.
TGNIIPI represents the test set of Gaussian noise and insufficient illumination polarization images.
TGNIICPI represents the test set of gaussian noise, insufficient illumination and clouds polarization images.

NFPI and TNFPI are mainly used for program debugging. GNIIPI and TGNIIPI corresponds to polarized skylight navigation in sunny weather. GNIICPI and TGNIICPI focuses on the influence of clouds on polarized skylight navigation.
*********************************************************************************************

*********************************************************************************************
%%%%%%%%   File name   %%%%%%%%%
Number_Yaw_Pitch_Roll_FocalLength_SolarAltitude_Turbidity_Wavelength_GroundAlbedo
unit: Yaw degree; Pitch degree; Roll degree; FocalLength mm; SolarAltitude degree; Turbidity none; Wavelength nm; GroundAlbedo none;

The first letter of the file name indicates the noise added to the polarization imger. The letter 'N' represents noise free polarization images (NFPI). The letter 'G' represents Gaussian noise and insufficient illumination polarization images (GNIIPI), the letter 'C' represents gaussian noise, insufficient illumination and clouds polarization images (GNIICPI). The letter 'TN' represents the test set of noise free polarization images (TNFPI). The letter 'TG' represents the test set of Gaussian noise and insufficient illumination polarization images (TGNIIPI). The letter 'TC' represents the test set of gaussian noise, insufficient illumination and clouds polarization images (GNIICPI).
*********************************************************************************************

*********************************************************************************************
%%%%%%%%   Parameters of Polarization Imager   %%%%%%%%%
The parameters of the polarization imager can refer to PHX050S-P Sony IMX250MZR COMS polarization imager, and the dataset is based on this polarization imager.
    Resolution is 2048*2448
    Pixel size is 3.45um
    Wavelength range is 380-780nm
    Format is 2/3"
    FocalLength is 4mm
*********************************************************************************************